\documentclass[12pt]{iopart}
\usepackage{iopams}  
\usepackage{graphicx}
\usepackage{color}
\begin{document}

\title[Critical nuclei]{A phase-field-crystal approach to critical nuclei}
\author{R. Backofen and A. Voigt}
\address{Institut f\"ur Wissenschaftliches Rechnen, TU Dresdem, 01062 Dresden, Germany}
\ead{axel.voigt@tu-dresden.de}

\begin{abstract}
We investigate a phase-field-crystal model for homogeneous nucleation. Instead of solving the
time evolution of a density field towards equilibrium we use a String Method to identify saddle
points in phase space. The saddle points allow to obtain the nucleation barrier and the critical 
nucleus. The advantage of using the phase-field-crystal model for this task is its ability to resolve
atomistic effects. The obtained results indicate different properties of the critical nucleus compared 
with bulk crystals and show a detailed description of the nucleation process.
\end{abstract}


\section{Introduction}

If a liquid is cooled below its melting temperature the liquid will exist in a metastable state until a nucleation event occurs. If the
source of nucleation in the undercooled melt is only due to fluctuation
phenomena the nucleation is called homogeneous. In classical nucleation theory
a spherical shape for the critical nuclei is assumed and its size is
determined as a result of competition between the bulk free energy reduction
and interfacial energy increase. If $V$ is the volume, $A$ the surface area,
$\Delta g$ the chemical free energy change per unit volume and $\gamma$ the
specific interfacial energy, the free energy change according to the formation
of a new phase is given by $\Delta G = V \Delta g + A \gamma$. For a
spherical shape with radius $r$ we thus obtain $\Delta G = 4/3 \pi r^3 \Delta
g + 4 \pi r^2 \gamma$. The radius $r^*$ of the critical nucleus must then be
such that $r^* = - 2 \gamma / \Delta g$ with the critical free energy of
formation of a critical nucleus $\Delta G^* = 16 \pi \gamma^3 / (3 (\Delta
g)^2)$. This classical theory has been utilized to interprete kinetics of many
phase transformations and have had some success for providing good
descriptions on the nucleation kinetics for various systems. On the other
hand, nucleation is generally significantly more complicated. The shape
might not be spherical due to an anisotropy of the interfacial energy between
a nucleus and the bulk phase, which results from the crystallographic nature
of a solid nuclei. Furthermore, the bulk properties of small nuclei may differ 
from bulk values typically obtained from larger samples. To account for these 
phenomena various new attempts in the context of diffuse interface models have 
been made to describe nucleation. Such a non-classical theory was pioneered by 
Cahn and Hilliard \cite{CahnHilliard_JCP_1959}. For subsequent studies, generalizations 
and specific applications to nucleation, we refer to
\cite{GranasyBoerzoenyyiPusztai_PRL_2002,GranasyPusztaiSaylorWarren_PRL_2007,ZhangChenDu_PRL_2007,WarrenPusztaiKoernyeiGranasy_PRB_2009} 
and the references therein. In these studies an order parameter is used to
distinguish between the nucleus and the bulk phase. Since nucleation takes
place by overcoming the minimum energy barrier, a critical nucleus is defined
as the order parameter fluctuation which has the minimum free energy increase
among all fluctuations which lead to nucleation. Therefore, the critical
nucleus can be found by computing the saddle points of the energy functional
of the order parameter, that has the highest energy in the minimum energy path
(MEP), which is the path whose highest energy is the lowest among all possible
paths. This is consistent with the large deviation theory which states that
the most probable path passes through the saddle point in the large time
limit. An efficient numerical approach for finding minimum energy paths and
saddle points, the so-called string methods (SM), has been introduced in
\cite{Eetal_PRB_2002}. The method is related to the nudged elastic band (NEB) method 
\cite{HenkelmanJonsson_JCP_2000}. Other approach are e.g. the minimax
method, which as been used in \cite{ZhangChenDu_JSC_2008} or 
a phase-field type approach, as used in \cite{Iwamatsu_JCP_2009} in the context of 
nucleation. We will here apply a simplified string method (SSM) \cite{Eetal_JChemP_2007} 
but not on an underlying diffuse-interface model but a more detailed phase-field-crystal
model \cite{Elderetal_PRL_2002}, which accounts for the discrete effects on the small length
scales involved in nucleation.

The outline of the paper is as follows: In Sec. 2 we introduce the
phase-field-crystal model as a local approximation of a classical dynamic
density functional theory. In Sec. 3 we describe the used string method. In Sec. 4 we
discuss implementational issues. In Sec. 5  we show results for homogeneous nucleation. 
Conclusions are drawn in Sec. 6.

\section{Phase-field-crystal model}

The phase-field-crystal (PFC) model is by now widely used in order to describe
solid-state phenomena on atomic length scales. The PFC model was first
developed in \cite{Elderetal_PRL_2002} and then subsequently applied to many
situations like interfaces
\cite{Athreyaetal_PRE_2007,BackofenVoigt_JPCM_2009}, polycrystalline pattern
formation \cite{Wuetal_PRB_2007, Goldenfeldetal_PRE_2005},
commensurate-incommensurate transitions \cite{Achimetal_PRE_2006}, edge
dislocations \cite{Berryetal_PRE_2006}, grain boundary pre-melting
\cite{Plappetal_PRB_2008}, colloidal solidification
\cite{vanTeeffelenetal_PRE_2009} and dislocation dynamics
\cite{BackofenBernalVoigt_IJMR_2010}. The model resolves the atomic-scale
density wave structure of a polycrystalline material and describes the
defect-mediated evolution of this structure on time scales orders of magnitude
larger than molecular dynamic (MD) simulations. In its simplest form the PFC model results from the energy 
\begin{equation}
\label{eq:0}
F[\varphi] = \int_\Omega - |\nabla \varphi|^2 + \frac{1}{2} (\Delta \varphi)^2 + f(\varphi) \; dx
\end{equation}
with $f(\varphi) = \frac{1}{2}(1 - \epsilon)\varphi^2 + \frac{1}{4} \varphi^4$ a potential, $\varphi$
the number density and $\epsilon$ a parameter determining the approximation of
the liquid structure factor ~\cite{Elderetal_PRL_2002}. Comparing
the energy with a classical phase-field type energy, e.g. $\int_\Omega
\frac{\delta}{2}|\nabla \phi|^2 + \frac{1}{\delta} g(\phi) \; dx$ for an order
parameter $\phi$, with $\delta$ a length scale determining the width of a
diffuse interface and $g(\phi)$ a double well potential, the difference is in
the sign of the gradient term and the additional higher order term. The
negative sign in the gradient term favors a changes in $\varphi$, whereas the
higher order term favors to suppress such changes. The competition between
both terms introduces a fixed length scale for which the energy will be
minimized. This length scale is used to model the periodicity of a crystal
lattice. The formulation used here favors a hexagonal closed packed structure 
in two dimensions. Due to the underlying periodicity, several solid state phenomena 
such as elasticity, plasticity, anisotropy and multiple grain orientations are 
naturally present in the formulation. The dynamic law constructed to minimize the 
free energy follows as the $H^{-1}$-gradient flow of the energy
\begin{equation}
\label{eq:1}
\partial_t \varphi = \Delta \mu
\end{equation}
with chemical potential $\mu = \frac{\delta F[\varphi]}{\delta \varphi}$ and the 
variational derivative given by 
\[
\frac{\delta F[\varphi]}{\delta \varphi} = \Delta^2 \varphi + 2 \Delta \varphi + f^\prime(\varphi).
\]
This defines the PFC model and its evolution is by construction towards a 
(meta) stable state.

Although this formulation is phenomenologically, the PFC model can be derived
starting from a Smoluchowski equation via dynamic density functional theory
using various approximations \cite{Elderetal_PRE_2007,vanTeeffelenetal_PRE_2009}. 
With an appropriate parameterization it thus provides also a quantitative atomic theory, 
operating on diffusive time scales. Within new developments 
\cite{vanTeeffelenetal_PRE_2009,Jaatinenetal_PRE_2009} quantitative agreement in 
computed properties could be achieved using the PFC model for various materials. 
It thus provides an ideal model to study nucleation.

\section{Minimum energy path}

\subsection{Definition}
\begin{figure}[htb]
\noindent
\center
\includegraphics*[angle = -0, width = 0.5 \textwidth ]{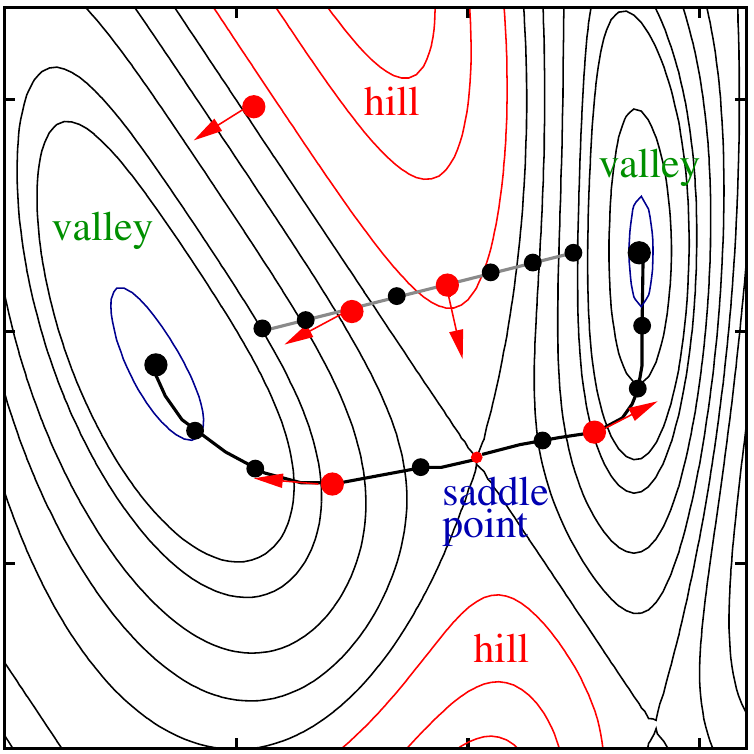}
\begin{center}
\begin{minipage}{0.9\textwidth}
\caption[short figure description]{
Schematic sketch of a free energy surface of a two dimensional phase
space. Two (meta) stable states (valleys) are separated by a saddle point. The
single states (circles) are evolved by thermodynamic forces (arrows). The
initial string (straight line) is evolved by the string method towards the MEP.  
\label{fig:1}
}
\end{minipage}
\end{center}
\end{figure}

The dynamics shown in eq. (\ref{eq:1}) describe the evolution towards equilibrium of a single state $\varphi$ in phase space according to 
the generalized thermodynamic force $\Delta \mu$. But in order to characterize nucleation the most likely transition path
between (meta) stable states has to be identified. In the description of such a non equilibrium process, the minimum energy path (MEP) plays a crucial
role. The MEP is a path in phase space that connects (meta) stable states. A path in phase space is thereby defined as 
\begin{eqnarray*}
\gamma_c = \left\{ \varphi_\alpha : \alpha \in [0,1] \right\} 
\end{eqnarray*}
with $\alpha$ a parameterization of the path. For the MEP the generalized thermodynamic force $\Delta \mu$ is tangential to this path:
\begin{eqnarray}
\left( \Delta \mu \right)^{\perp} =0
\end{eqnarray}
Thus, using the dynamics in eq. (\ref{eq:1}), any state of the MEP will always evolve along the MEP towards a stable state. That is, the MEP is a real 
reaction path in phase space and along the MEP the energy is defined by eq. (\ref{eq:0}). The local energetic maxima and minima along the MEP can be used to
determine the nucleation barrier and the critical nucleus. The string method (SM) is been 
designed to find the MEP. It evolves a given chain of states towards the MEP, see Fig. \ref{fig:1} for illustration. 

\subsection{String method}
A path in phase space $\gamma_c$ is represented by a discrete set of states $\varphi_i$, denoted by
\begin{eqnarray*}
\gamma = \left\{ \varphi_i : i = 0,1, \ldots, N \right\} = \left\{  \varphi_i \right\}.
\end{eqnarray*}
The length of the path is defined by
\begin{eqnarray*}
L(\gamma)=\sum_{i=0}^{N-1} \left| \varphi_{i+1}-\varphi_i \right| 
\quad {\rm and } \quad
\left| \varphi \right| = \int \left| \varphi(r) \right| \,dr
\end{eqnarray*}
where $|\cdot|$ measures the distance between two states and is defined by the $L^2$-norm. Thus,
the normalized tangent $\hat{t}$ to the path at state $\varphi_i$ may be calculated as 
$\hat{t_i}=\frac{\varphi_{i+1}-\varphi_i}{\left|\varphi_{i+1}-\varphi_i\right|}$.

The idea behind the SM is to minimize the free energy of all single states
according to 
\begin{eqnarray}
\varphi_i^{n+1}=\tau \Delta \mu + \varphi_i^n.
\end{eqnarray}
but restricting the evolution to orthonormal direction of the path. In addition a tangential
force is included in order to keep the quality of the representation of the
path by the string. 
\begin{eqnarray} \label{eq-SM}
\gamma^n \rightarrow& \gamma^{n+1}&= \left\{ \varphi_i^{n+1} \right\} \\ \nonumber 
{\rm with}&\varphi_i^{n+1} &=\tau \left( \Delta \mu \right)^\perp+ \varphi_i^n
+\lambda_i \hat{t}
\end{eqnarray}
The Langrang multipliers $\lambda_i$ are e.g. uniquely determined by forcing 
an equidistant distribution of states along the path. $\tau$ is a fictitious
timestep and controls the velocity of evolution. Eq. (\ref{eq-SM}) defines the 
String Method. It is easily seen that the MEP $\gamma_{\rm MEP}$ is an invariant 
according to the dynamics of SM. By definition of the MEP, the thermodynamic
force  is only tangential to the path. The constraints introduced by the
tangential force do not alter the path, but only reparameterize the representation of $\gamma$. Thus, $\gamma_{\rm MEP}^n$ and $\gamma_{\rm MEP}^{n+1}$ represents the same path 
in phase space.

In order to implement the SM the thermodynamic force has to be calculated and
projected to the orthogonal direction of the path. Furthermore the Langrange
multipliers have to be calculated. In order to simplify the calculation the SM
can be divided in two steps leading to a Simplified String Method (SSM). First 
the string is evolved due to the
thermodynamic force and then the path is reparameterized, see \cite{Eetal_JChemP_2007}.
That is, the states representing the path $\left\{\tilde{\varphi}_i\right\}$ are replaced by 
equally distant states, that represent the same path $\left\{\varphi_i\right\}$. 
This new states are constructed by interpolation between the original states 
$\left\{\tilde{\varphi_i}\right\}$. As in every evolution step there is
a parameterization step, it is no longer necessary to project the thermodynamic force.

Thus the SSM is defined by two steps:
\begin{enumerate}  
\item Evolution step:
\begin{eqnarray}
\gamma^n \rightarrow& \tilde{\gamma}^{n+1}&= \left\{ \tilde{\varphi}_i^{n+1} \right\} \\ \nonumber 
{\rm with}&\tilde{\varphi}_i^{n+1} &= \tau \Delta \mu  + \varphi_i^n
\end{eqnarray}
\item Reparameterization step:
\begin{eqnarray}
&\tilde{\gamma}^{n+1} \rightarrow \gamma^{n+1}=\left\{\varphi_i^{n+1}\right\}\\ \nonumber 
{\rm such~that}& 
\left| \varphi_{i+1}^{n+1}-\varphi_i^{n+1} \right| =
 \frac{L(\tilde{\gamma}^{n+1})}{N-1} \quad i = 0,1,2, \ldots,N-1
\end{eqnarray}
and $\gamma^{n+1}$ and $\tilde{\gamma}^{n+1}$ representing the same path in phase space. 
\end{enumerate}
Here the reparameterization is done to force equidistant states on the
path. However, the reparameterization may also be changed to account for
problem specific details, e.g. to get finer representation at the saddle point.
The advantage of SSM over SM is that the thermodynamic force has not to be
projected. Additionally it is shown that this modification leads to a more stable 
and accurate method \cite{Eetal_JChemP_2007}.

\subsection{Fixed length Simplified String Method}
In the above defined SSM an initial path in phase space is evolving
towards the MEP. The first and the last state representing the path thereby converge to
different (meta) stable states. The saddle point or here the critical nucleus is defined
by the state of highest energy in the MEP. If there is only one energetic
maximum and the saddle point is well defined, the MEP could be calculated easily by
just solving the time evolution of a small perturbation of the critical nucleus towards 
the stable states according to eq. (\ref{eq:1}). Thus, only two time dependent simulations 
have to be done. Therefore, it is
enough to find the saddle point to reconstruct the whole MEP efficiently.
In order to concentrate the simulation effort to find only the saddle point, we 
introduce a Fixed Length Simplified String Method (FLSSM). The
total length of the string is restricted by the reparameterization step. Assume
that we allow a maximal length of the string, $L_{\rm fixed}$. For simplicity we also
assume that the first state converges towards a stable state. Then, the
reparameterization can always project the states back to a string beginning
with the first state with length $L_{\rm fixed}$. The last state is not converging to a meta stable state, but
might be some unstable state but within a different basin. This state can be used to reconstruct the whole MEP by
a time evolution according to eq. (\ref{eq:1}). The same idea can be used at the
same time on both sides of the saddle point, by fixing a state near the saddle
point and restricting the length on both sides of the path. As we need some
information about the string length and the position of the saddle point, we use 
the FLSSM in order to refine and improve accuracy of a MEP which was 
calculated by the standard SSM with only a few states. The method can be viewed as an adaptive approach
which efficiently finds the saddle point within a given tolerance.

\subsection{Implementation of a Fixed Length Simplified String Method for PFC}
The string is a set of density distribution $\left\{\varphi_i\right\}$. As we
consider a closed system and have mass conserving dynamics, we have to
restrict the possible states representing a string to the same mean density,
$\bar{\varphi} = \int \varphi_i(r) \, dr$ for all $i$. 

For every state the standard dynamics has to be solved according to eq. (\ref{eq:1}).
The partial differential equation of 6th order is splitted into a set of three
second order equations:
\begin{eqnarray*}
\partial_{t} \varphi_i &=& \Delta \mu \\
\mu &=& \Delta v + 2 \Delta \varphi_i + f^\prime(\varphi_i) \\
v &=& \Delta \varphi_i
\end{eqnarray*} 
for which a stable semi-implicit finite element discretization is introduced
in \cite{Backofenetal_PM_2007}. We use this approach but with higher order
elements. The algorithm is implemented
in the adaptive finite element toolbox AMDiS \cite{Veyetal_CVS_2007}. 

The fictitious timestep is adjusted such that the reparameterization step 
can be done mostly considering only neighbouring states and such that the 
evolved state is substantial different from the previous one.    

In this work we use linear interpolation between the states. We define the
length of the string up to state M in analogy to $L(\gamma)$ for the whole string by
$L_M(\gamma)=  \sum_{i=0}^{M-1} \left|\varphi_{i+1}-\varphi_i \right|$. The distance
between states after reparameterization is $\bar{l}=\frac{L(\tilde{\gamma})}{N-1}$. 
Then the reparameterized state $\varphi_i$ at $L^*=i\bar{l}$ is constructed 
by linear interpolation using the neighboring states of $\varphi_i$ from $\tilde{\gamma}$.
\begin{eqnarray}
\varphi_i=
\tilde{\varphi}_k+(\tilde{\varphi}_{k+1}-\tilde{\varphi}_{k})\alpha \\
\alpha = \frac{L^*-L_k(\tilde{\gamma})}{\left|\tilde{\varphi}_{k+1}-\tilde{\varphi}_k \right|}
\quad {\rm and} \quad L_k(\tilde{\gamma}) \leq L^*  < L_{k+1}(\tilde{\gamma})
\end{eqnarray}

The FLSSM is implemented in a parallel way. That is, every state define a
process and the evolution step is calculated in parallel. The result is then
send to the nodes of the neighbouring states. In order to avoid complicated
communication between the processes, the reparameterization step is further 
simplified. The linear interpolation described above is used if the reparameterized
state is in between the neighbouring states. If not, the reparameterized state is
set to one of the neighbouring sites.
\begin{eqnarray}
\varphi_i= \left\{ 
\begin{array}{ll}
\varphi_{i-1},&  L^*  < L_{i-1}(\tilde{\gamma}) \\
{\rm linear~interpolation},& L_{i-1}(\tilde{\gamma}) \leq L^*  \leq 
  L_{i+1}(\tilde{\gamma}) \\
\varphi_{i+1},&  L^*  > L_{i+1}(\tilde{\gamma})
\end{array}
\right.
\end{eqnarray}
This does not alter the MEP but only the dynamic of the string in phase
space, as long as we choose the fictitious timestep in a way that for pathes near
the MEP at the end of simulation always linear interpolation between neighbouring sites can be used. 

We define convergence of the method if the string changes only in tangential direction. This is ensured by two criteria. First, the change in string length 
and the change in energy in every state after reparameterization is below a given tolerance which ensures no change of the string in normal direction. The second criteria accounts for the change in tangential direction and ensures the evolution towards the (meta) stable states. Therefore the fictitious time step is adjusted to evolves the states substantially far but still allowing for linear interpolation with $0.3 < \alpha < 0.7$. If the second criteria is not fullfilled, the fictitious time step is adapted. Due to small thermodynamic forces near the saddle point, the second criteria may be relaxed for states very close to the saddle point.

\section{Results}

We consider as a proof of concept the nucleation of a crystal grain in an undercooled liquid. The parameters needed in the PFC model, eq. (\ref{eq:1}), are the mean value of the density, $\bar{\varphi}=\int \varphi(r) \, dr$ and the parameter $\epsilon$, which can be interpreted
as a driving force of the phase change, e.g. undercooling \cite{Majaniemietal_PRE_2009,Yuetal_2010} or
strength of interaction \cite{vanTeeffelenetal_PRE_2009}. In our example we choose parameters in the coexistence region of the phase diagram
$(\epsilon, \bar{\varphi}) = (-0.289, -0.345)$. For this parameter the liquid is a meta stable state. The stable state is a grain in coexistence with liquid. The grain is slightly anisotropic and there is a small density difference between crystal and liquid, see \cite{BackofenVoigt_JPCM_2009}. 

\begin{figure}[htb]
\noindent
\center
\includegraphics*[angle = -0, width = 0.7 \textwidth ]{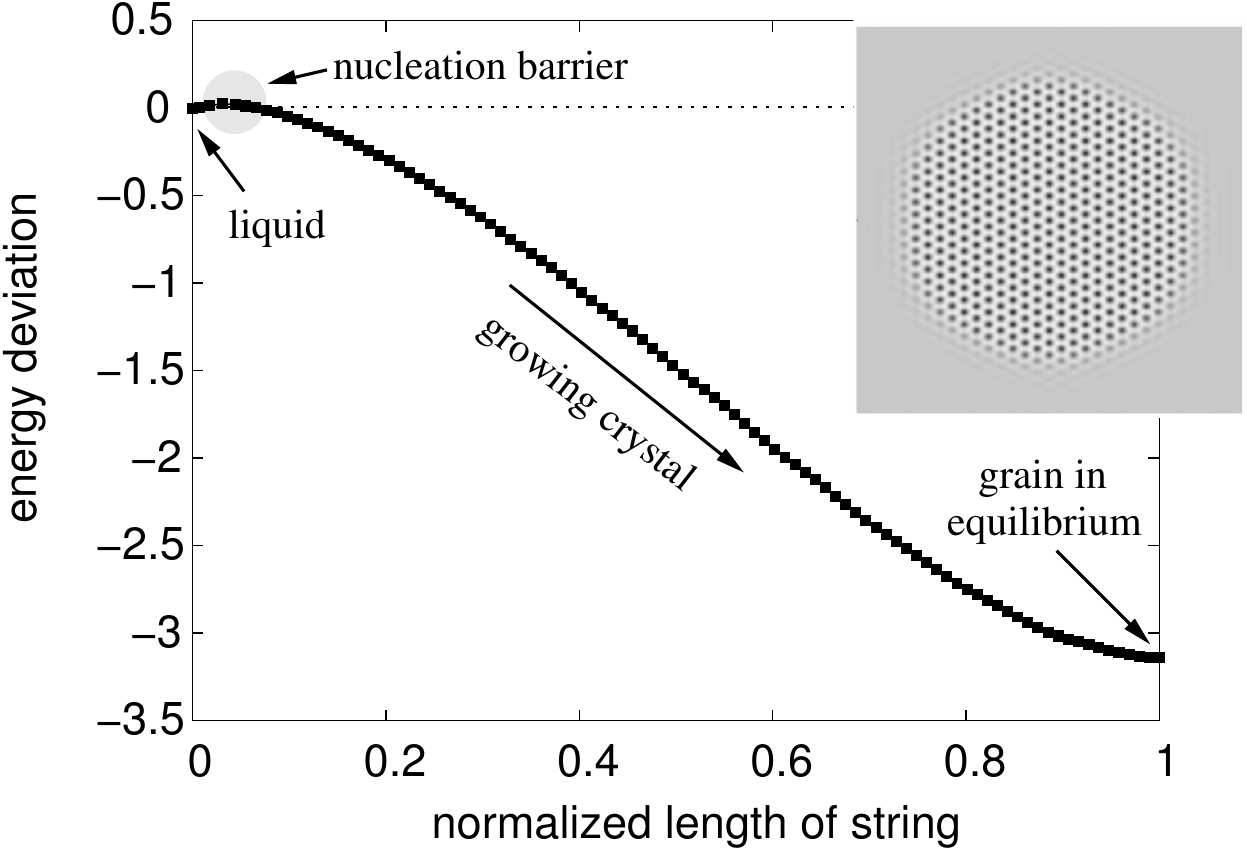}
\begin{center}
\begin{minipage}{0.9\textwidth}
\caption[short figure description]{
Free energy along the MEP. The free energy is plotted relative to the liquid
state at normalized string length, l=0. The path is discretized by 94
states. At l=1 the stable state is reached.  
\label{fig-MEP}
}
\end{minipage}
\end{center}
\end{figure}

In order to calculated the MEP an initial string $\gamma^0=\left\{\varphi_i^0\right\}$ has to
be defined such that the mean density of every state is
equal $\bar{\varphi}=\bar{\varphi}_i^0$ and that the first and the last state
evolve towards two different (meta) stable states. In our work, we use two different
initial strings to demonstrate that the obtained MEP is independent of the initial configuration.
In the first example we set the first state to liquid $\varphi_0(x)=\bar{\varphi}$ and the last state 
to the equilibrium shape of the grain. The states in between are then constructed by linear
interpolation. In the second example, every state was set homogeneous and disturbed by
white noise $\eta$ such that $\varphi_i(x)=\bar{\varphi}+S_i \eta$. The strength of the
noise $S_i$, was linearly increased starting from $0$ to represent the liquid state towards a large value, which ensures evolution within the coexistence regime. 
Both initial strings converge to the same MEP shown in Fig.~\ref{fig-MEP}. In order to proof
stability of the obtained MEP every state was disturbed independently by some random
field and than taken as a initial string to recalculate the MEP.    

\begin{figure}[htb]
\noindent
\center
\includegraphics*[angle = -0, width = 0.7 \textwidth ]{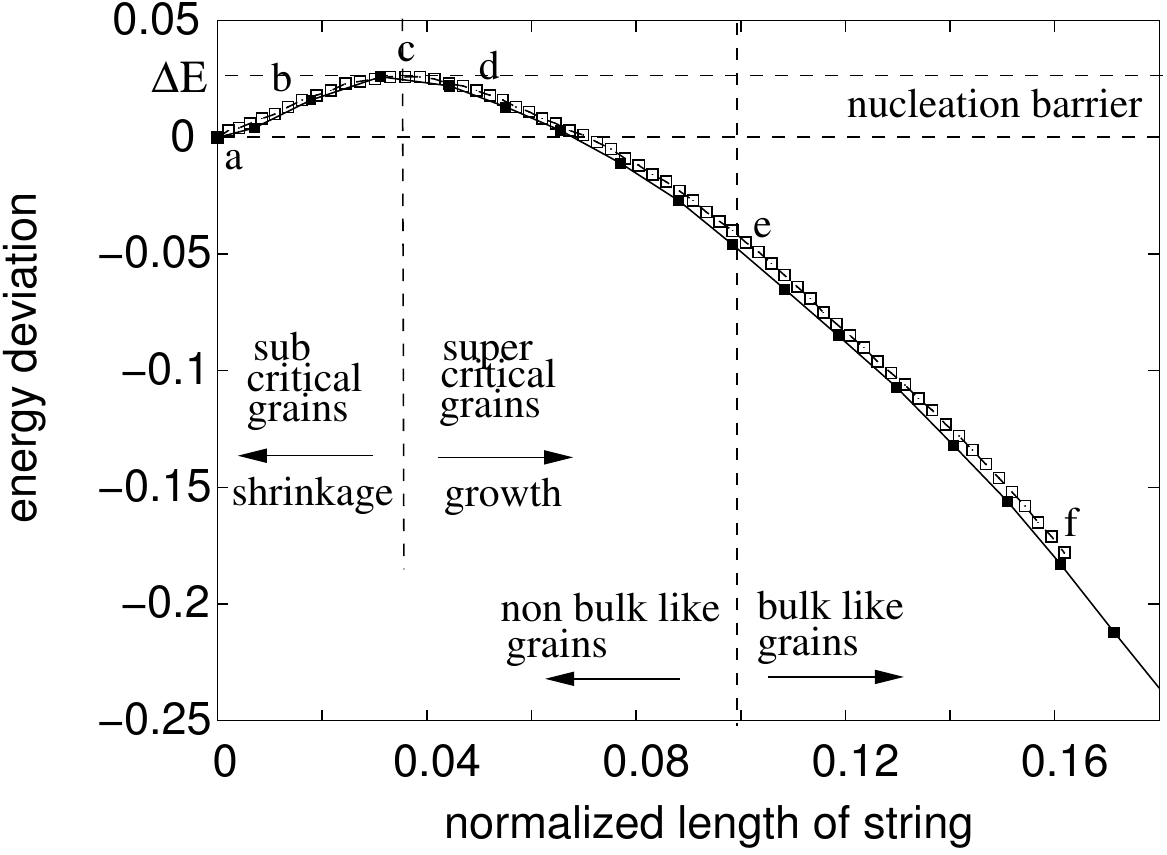}
\begin{center}
\begin{minipage}{0.9\textwidth}
\caption[short figure description]{
Detailed MEP around the saddle-point. Closed symbols indicate the MEP
calculated by SSM as in Fig. \ref{fig-MEP}. Open symbols show the result achieved by
restricting the string length to $\frac{1}{6}$ and using FLSSM. 
\label{fig-MEP2}
}
\end{minipage}
\end{center}
\end{figure}

The first state corresponds to liquid and the last to a grain in coexistence with the 
liquid. The grain equilibrium
is energetically favourable compared to the liquid and is the stable state in
phase space. The liquid state is meta stable. The nucleation barrier or
the saddle point is found at normalized string length of approx. $0.04$. States
right to the saddle point correspond to growing crystallites and left to melting
crystallites. The string was
discretized by 94 states which are equally distributed, so the region around the nucleation 
barrier is resolved only by 10 states. In order to get a better resolution the FLSSM is used. 
The length of the string is therefore restricted to $\frac{1}{6}$ of the original length of 
the MEP and is rediscretized by 46 states, which are constructed by linear interpolation of the
calculated MEP. We can view this as an adaptive method to increase the accuracy of the calculated
saddle point or a proof that the obtained saddle point is independent of the used parameterization
of the string. In our example this independency is shown. Fig. \ref{fig-MEP2} shows the obtained nucleation barrier 
$\Delta E$ and critical nucleus. 
      
The critical nucleus is defined by the state indicated by (c), 
$\varphi_{\rm c}$. (b) indicates $\varphi_{\rm b}$ a sub critical state, which
most likely will melt. (d) - (f) indicate states $\varphi_{\rm d}$ - $\varphi_{\rm f}$ which correspond to
super critical states which will solidify. 
 
\begin{figure}[htb]
\noindent
\center
\begin{tabular}{cccc}
b & \includegraphics[width = 0.2 \textwidth ]{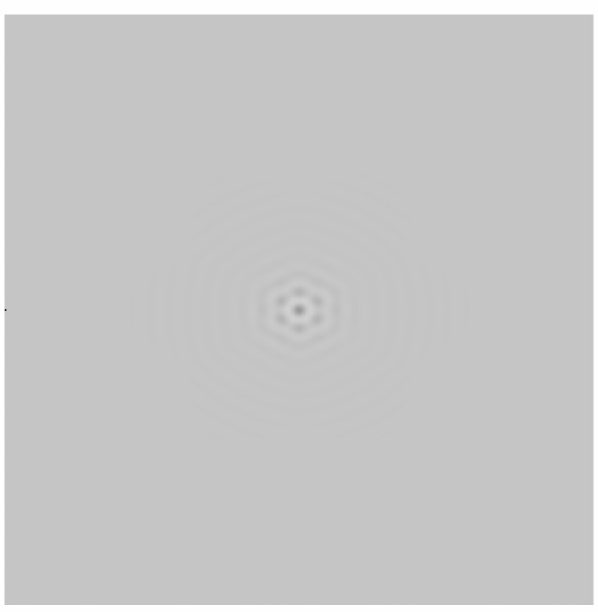} &
c & \includegraphics[width = 0.2 \textwidth ]{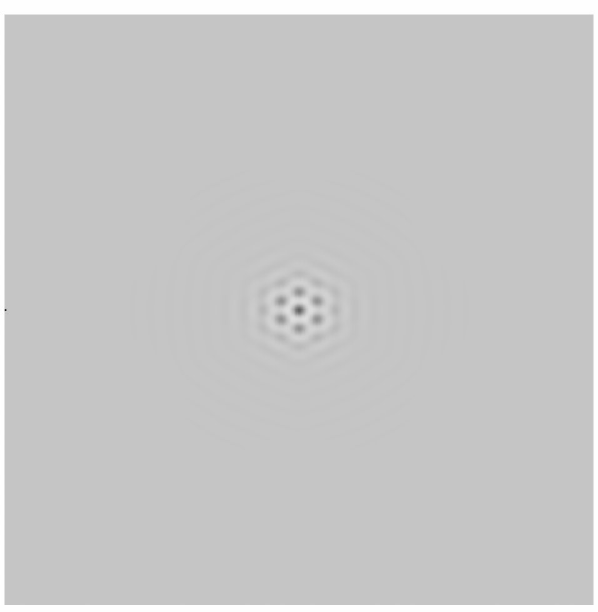} \\
d & \includegraphics[width = 0.2 \textwidth ]{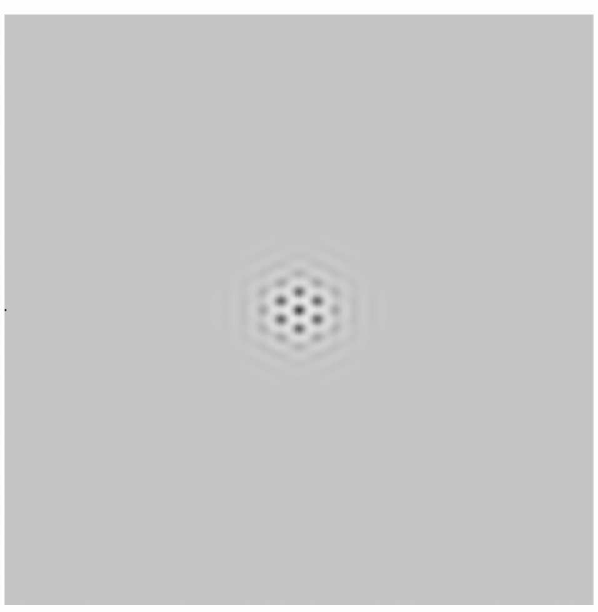} &
e & \includegraphics[width = 0.2 \textwidth ]{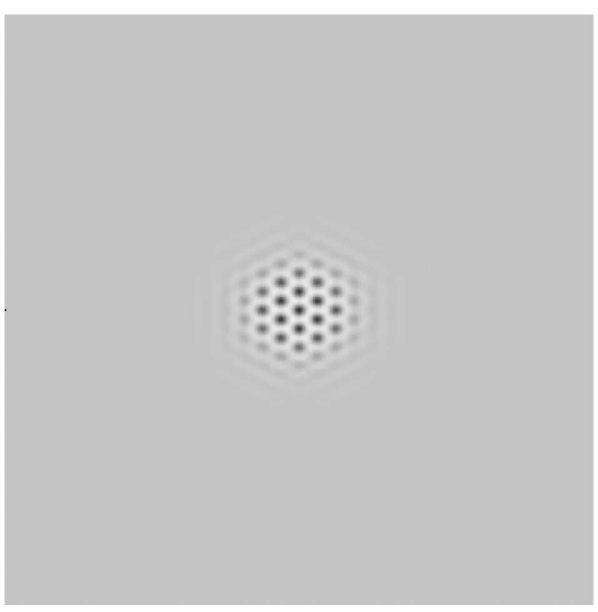}\\
f & \includegraphics[width = 0.2 \textwidth ]{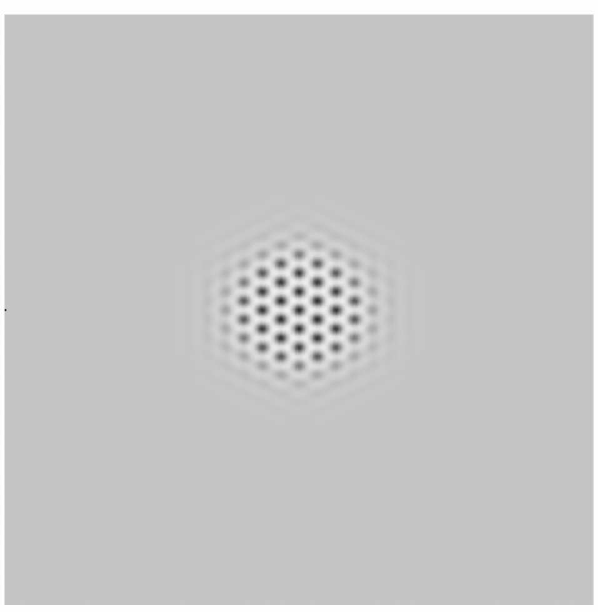} &
  & \includegraphics[width = 0.2 \textwidth ]{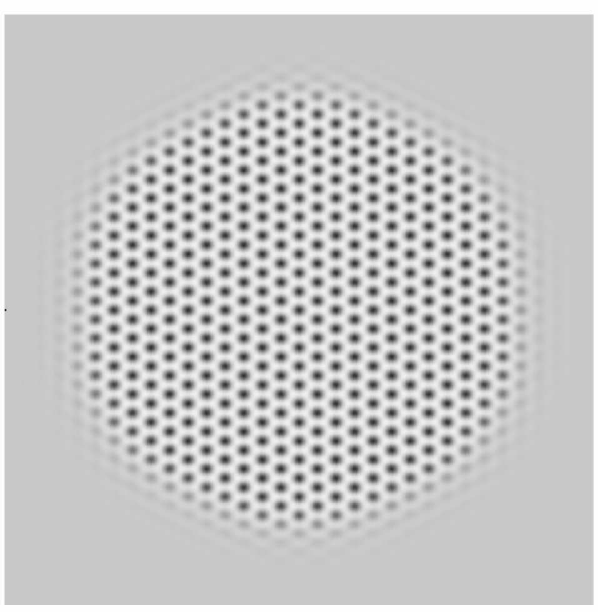}  
\end{tabular}
\begin{center}
\begin{minipage}{0.9\textwidth}
\caption[short figure description]{
Various states at the MEP. The labels (b)-(f) indicate the different state $\varphi_{\rm b}$ - $\varphi_{\rm f}$. 
\label{fig-grain}
}
\end{minipage}
\end{center}
\end{figure}

\begin{figure}[htb]
\noindent
\center 
\includegraphics*[angle = 0, width = 0.7 \textwidth ]{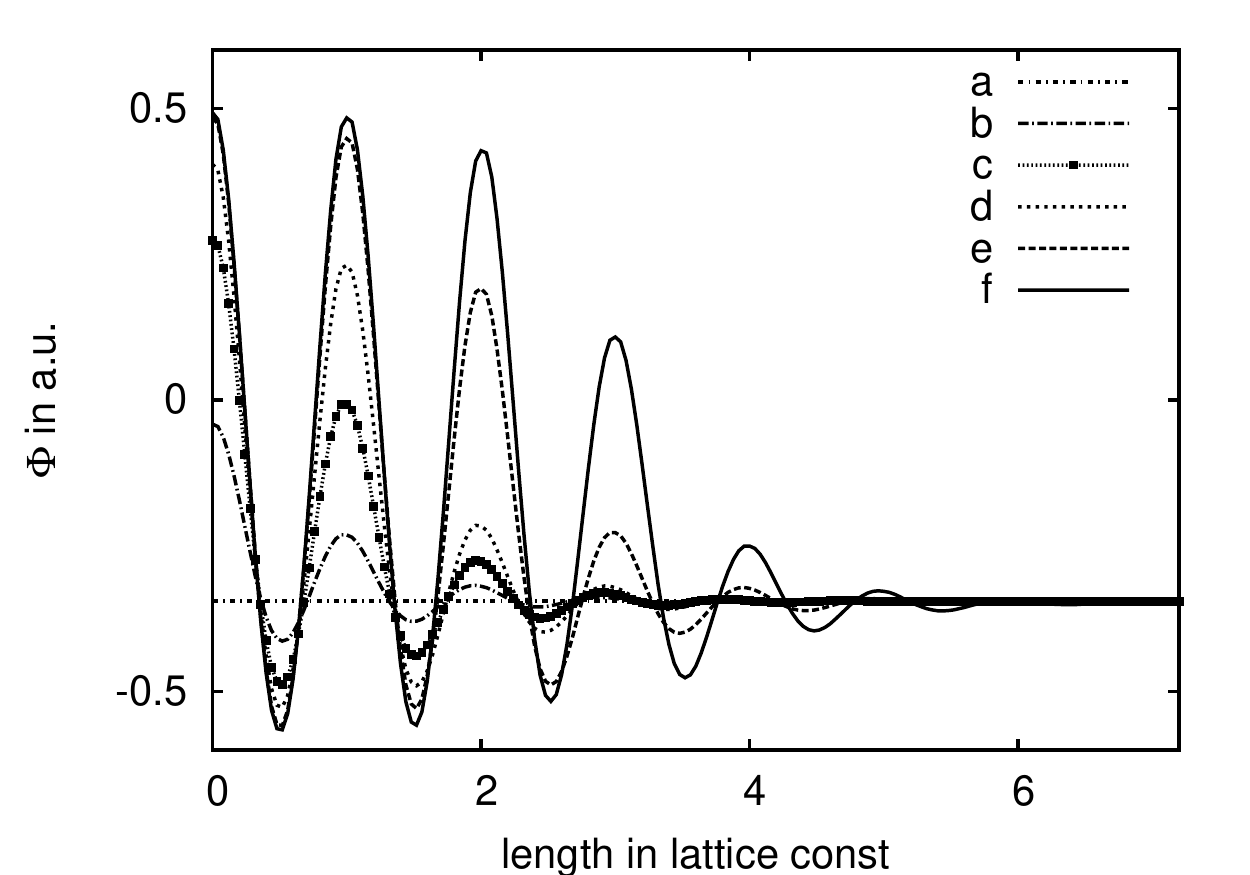}
\begin{center}
\begin{minipage}{0.9\textwidth}
\caption[short figure description]{ 
Density profile of grains at the MEP. The labels (b)-(f) indicate the
different state $\varphi_{\rm a}$ - $\varphi_{\rm f}$, see~Fig.~\ref{fig-MEP2}.
\label{fig-prof}
}
\end{minipage}
\end{center}
\end{figure}

In Fig.~\ref{fig-grain} the density field of the labeled states are
shown. The critical nucleus is a hexagonal cluster with only seven maxima. A small perturbation of this state will lead either to growth towards the equilibrium shape
or to melting. The grow is symmetric and can be seen more quantitative in Fig.~\ref{fig-prof}, which shows the density profile along the x-axis in the various states
of the growth process. The density plot shows that the maximum amplitude of the critical nucleus is smaller than in the final bulk state. This can correspond
to defects in the crystal, as we consider here only a mean-field description, or weaker ordering of particles. In both cases this shows that the critical
nucleus has different structure and bulk energy than the corresponding bulk state. Nucleation thus begins with a disturbance that reflects the crystal
structure but has a small amplitude. During growth the spatial size of the initial fluctuation and the amplitude increases. At state $\varphi_{\rm e}$ the maximum amplitude is equal to the bulk value and does not increase anymore. After this state the grain begins to grow only along the solid liquid phase boundary.   
    
These computations indicate that the subcritical and supercritical grains around the critical nucleus are different to ideal bulk crystals.
In addition to the non-spherical shape of the nucleus, which is not considered in classical nucleation theory, such size dependent properties 
are also not considered in full detail in classical phase field approaches for nucleation. 

\section{Conclusion}

A phase-field-crystal model is used to determine nucleation barriers and the critical nucleus in homogeneous nucleation. The results obtained
indicate details of the nucleation process which are not considered in classical nucleation theory but also cannot be addressed in full detail 
with classical phase field models. The obtained size of the critical nucleus, which here consists only of 7 atoms furthermore asks for an 
atomistic description. Even if only phenomenological values are used in the computation, the described method gives a proof of concept. 
The described String Method is independent of the parameterization of the underlying evolution model and thus will allow also to be used for
specific materials. With the described implementational details of the String Method, concerning parallel processing and adaptive concepts we believe 
the approach to be applicable also in three dimensions.

\vspace*{0.5cm}
{\bf Acknowledgments} We would like to thank E. Vanden-Eijnden for valuable
discussions. RB and AV acknowledges support of the DFG via Vo899/7-2 within
SPP 1296.


\section*{References}
\begin{thebibliography}{12}

\bibitem{CahnHilliard_JCP_1959}
J.W.~Cahn and J.E.~Hilliard,{Free energy of a nonuniform System. 3. Nucleation in a 2-component incompressible fluid.3}, J. Chem. Phys. {\bf 31} (1959) 688-699

\bibitem{GranasyBoerzoenyyiPusztai_PRL_2002}
L.~Gr\'an\'asy, T.~B\"orzs\"onyi and T.~Pusztai, {\em Nucleation and Bulk Crystallization in Binary Phase Field Theory}, Phys. Rev. Lett. {\bf 88} 206105 

\bibitem{GranasyPusztaiSaylorWarren_PRL_2007}
L.~Gr\'an\'asy, T.~Pusztai, D.~Saylor and J.A.~Warren, 
{\em Phase Field Theory of Heterogeneous Crystal Nucleation}, Phys. Rev. Lett. {\bf 98} (2007) 035703

\bibitem{ZhangChenDu_PRL_2007}
L.~Zhang, L.-Q.~Chen and Q.~Du, {\em Morphology of Critical Nuclei in Solid-State Phase Transformations}, Phys. Rev. Lett. {\bf 98} (2007) 265703

\bibitem{WarrenPusztaiKoernyeiGranasy_PRB_2009}
J.A.~Warren, T.~Pusztai, L.~K\"ornyei and L.~Gr\'an\'asy, {\em Phase field approach to heterogeneous crystal nucleation in alloys}, Phys. Rev. B {\bf 79} (2009) 014204 

\bibitem{Eetal_PRB_2002}
W.~E, W.~Ren and E.~Vanden-Eijnden,{ \em String method for the study of rare
  events}, Phys. Rev. B {\bf 66} (2002) 052301

\bibitem{HenkelmanJonsson_JCP_2000}
G. Henkelman and H. Jonsson, {\em Improved tangent estimate in the nudged elastic band method for finding minimum energy paths and saddle points}, J. Chem. Phys. {\bf 113} (2000) 9978-9985

\bibitem{ZhangChenDu_JSC_2008}
L.~Zhang, L.-Q.~Chen and Q.~Du, {\em Mathematical and Numerical Aspects of a Phase-field Approach to Critical Nuclei Morphology in Solids}, J. of Sci. Comp. {\bf 37} (2008) 89-102

\bibitem{Iwamatsu_JCP_2009}
M.~Iwamatsu, {\em Minimum free-energy path of homogenous nucleation from the phase-field equation}, J. Chem. Phys. {\bf 130} (2009) 244507

\bibitem{Eetal_JChemP_2007}
W.~E, W.~Ren and E.~Vanden-Eijnden,{\em Simplified and improved string method
  for computing the minimum energy paths in barrier-crossing events},
J. Chem. Phys. {\bf 126} (2007) 164103

\bibitem{Elderetal_PRL_2002}
K.R. Elder, M. Katakowski, M. Haataja and M. Grant, {\em Modeling elasticity
  in crystal growth}, Phys. Rev. Lett. {\bf 88} (2002) 245701 

\bibitem{Athreyaetal_PRE_2007}
B.P. Athreya, N. Goldenfeld, J.A. Danzig, M. Greenwood and N. Provatas, {\em
  Adaptive mesh computation of polycrystalline pattern formation using a
  renormalization-group reduction of the phase-field crystal model},
Phys. Rev. E {\bf 76} (2007) 056706 

\bibitem{BackofenVoigt_JPCM_2009}
R. Backofen, and A. Voigt, {\em Solid-liquid interfacial energies and
  equilibrium shapes of nanocrystals}, J. Phys. Cond. Mat. {\bf 21} (2009) 464109 

\bibitem{Wuetal_PRB_2007}
K.-A. Wu and A. Karma, {\em Phase-field crystal modeling of of equilibrium
  bcc-liquid interfaces}, Phys. Rev. B {\bf 76} (2007) 184107 

\bibitem{Goldenfeldetal_PRE_2005}
N. Goldenfeld, B.P. Athrya and J.A. Dantzig, {\em Renormalization group
  approach to multiscale simulation of polycrystalline materials using the
  phase-field crystal model}, Phys. Rev. E {\bf 72} 020601 

\bibitem{Achimetal_PRE_2006}
C.V. Achim, M. Karttunen, K.R. Elder, T. Ala-Nissil\"a and S.C. Ying, {\em
  Phase diagram and commensurate-incommensurate transitions in the phase-field
  crystal model wwith an externam pinning potential}, Phys. Rev. E {\bf 74}
(2006) 021104 
 
\bibitem{Berryetal_PRE_2006}
J. Berry, M. Grant and K.R. Elder, {\em Diffuse atomic dynamics of edge
  dislocations in two dimensions}, Phys. Rev. E {\bf 73} (2006) 031609 

\bibitem{Plappetal_PRB_2008}
J. Mellenthin, A. Karma and M. Plapp, {\em Phase-field crystal study of
  grain-boundary premelting}, Phys. Rev. B {\bf 78} (2008) 184110 

\bibitem{vanTeeffelenetal_PRE_2009}
S. van Teeffelen, R. Backofen, A. Voigt and H. L\"owen, {\em Derivation of the
  phase field crystal model for colloidal solidification} 
Phys. Rev. E {\bf 79} (2009) 051404

\bibitem{BackofenBernalVoigt_IJMR_2010}
R. Backofen, F. Bernal, A. Voigt, {\em Elastic interactions in phase-field-crystal models - numerics and postprocessing.}, Internat. Journal of Materials Research (accepted)

\bibitem{Elderetal_PRE_2007}
K.R. Elder, N. Provatas, J. Berry, P. Stefanovic and M. Grant, {\em
  Phase-field crystal modeling and classical density functional theory of
  freezing}, Phys. Rev. B {\bf 75} (2007) 064107 

\bibitem{Backofenetal_PM_2007}
R. Backofen, A. R\"atz and A. Voigt, {\em Nucleation and growth by a
  phase-field crystal (PFC) model}, Phil. Mag. Lett. {\bf 87} (2007) 813-820 

\bibitem{Jaatinenetal_PRE_2009}
A.~Jaatinen, C.V.~Achim, K.R.~Elder and T.~Ala-Nissila, {Thermodynamics of bcc metals in phase-field-crystal models}  Phys. Rev. E {\bf 80} (2009) 031602

\bibitem{Veyetal_CVS_2007}
S. Vey and A. Voigt, {\em AMDiS - Adaptive multidimensional simulations},
Comput. Vis. Sci. {\bf 10} (2007) 57-66 

\bibitem{Yuetal_2010}
Y.-M. Yu, R. Backofen, A.Voigt, {Modelling heteroepitaxial growth of thin films on vicinial substrates using phase-filed-crystal approach.} (submitted)

\bibitem{Majaniemietal_PRE_2009}
S. Majaniemi and N. Provatas, {\em Deriving surface-energy anisotropy for
  phenomenological phase-field models of solidification}, Phys. Rev. E {\bf
  79} (2009) 011608 

\end{thebibliography}
\end{document}